KEYNOTE PAPER

# CELL-SURFACE ADHESIVE INTERACTIONS IN MICROCHANNELS AND MICROVESSELS

Michael R. King, mike_king@urmc.rochester.edu

**Department of Biomedical Engineering**
University of Rochester
601 Elmwood Ave., Box 639
Rochester, New York 14642 USA

**ABSTRACT**

Adhesive interactions between white blood cells and the interior surface of the blood vessels they contact is important in inflammation and in the progression of heart disease. Parallel-plate microchannels have been useful in characterizing the strength of these interactions, in conditions that are much simplified over the complex environment these cells experience in the body. Recent computational and experimental work by several laboratories have attempted to bridge this gap between behavior observed in flow chamber experiments, and cell-surface interactions observed in the microvessels of anesthetized animals.

**INTRODUCTION**

The interplay between transient chemical bonding and hydrodynamic forces in the microcirculation controls the eventual fate of blood cells. Spherical white blood cells (neutrophils) reversibly bind to the interior surface of venules through a family of adhesion molecules known as selectins and translate at an average velocity much lower (0.1-50%) than that of freely suspended particles near the wall (Lawrence and Springer, 1991). This enables the bound neutrophil to communicate with the layer of endothelial cells that line venules through a cell signaling cascade (Ebnet and Vestweber, 1999), at times causing the neutrophil to firmly deposit at a site of injury as needed. Disruption of the normal function can lead to inflammatory diseases or interfere with lymphocyte/stem cell homing (Springer, 1995). Obtaining an accurate picture of cell rolling in vivo is complicated by the large number of chemical species involved, with at least three selectin molecules and five nonexclusive binding partners, each exhibiting quantitative differences in binding kinetics (Ebnet and Vestweber, 1999). Furthermore, physical parameters are constantly varying such as the wall shear rate and the local surface coverage of adhesion molecules as cells become activated or relax to baseline levels. These complexities have been effectively dealt with by recreating the rolling phenomena in a cell-free system, replacing the endothelium with an immobilized layer of a single adhesion molecule and perfusing this surface with either neutrophils (Goetz et al., 1994) or carbohydrate-coated spherical beads (Brunk et al., 1996). Computational models have also been quite effective at integrating information such as kinetic rate constants and surface compliance derived in idealized experimental systems, and predicting the resulting nonlinear behavior in concentrated cell suspensions encountered in vivo.

Biologists have identified many of the molecular constituents that mediate adhesive interactions between white blood cells, the cell layer that lines blood vessels, blood components, and foreign bodies. However, the mechanics of how blood cells interact with one another and with biological or synthetic surfaces is quite complex: owing to the deformability of cells, the variation in vessel geometry with diameters ranging from $O(10^{-6}\text{-}10^{-2})$ m, and the large number of competing chemistries present (Lipowski et al., 1991, 1996). There have been few attempts to model cell adhesion between that of considering a single isolated cell and large scale continuum approaches that do not account for the finite size (and resulting physical interactions) of suspended or surface-bound cells. Finite element analyses of the flow in large arteries, approximating the blood suspension as a single phase with bulk rheological properties of a viscoelastic fluid, have been useful in identifying areas most prone to the development of atherosclerotic plaques due to flow reversal and recirculating eddies (Hyun et al., 2000). However, progression of atherosclerosis and acute cases of platelet-mediated thrombosis are controlled by the adhesion of cells to the vessel wall. Close to the wall the finite size effects of blood cells become important and their neglect will remain a barrier to accurate models of the accumulation rate of adherent cells on the surface of vessels or surgical implants (King and Leighton, 1997).

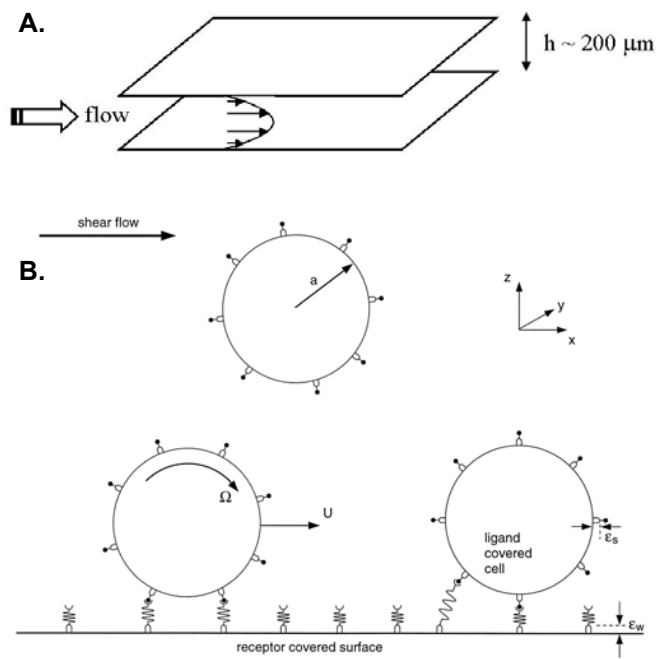

**Figure 1. The experimental and theoretical systems. A. The flow chamber consists of two parallel plates, with the lower plate functionalized with adhesion molecules. B. Schematic of MAD: a rigorous calculation of suspension flow near a wall is coupled to a stochastic model of specific surface adhesion through reversible receptor-ligand bonds.**

Another example of a physiological flow where the effects of finite cell size are important is in capillaries that can be smaller in diameter than a resting leukocyte (Bathe et al., 2002). Adhesive cells or leukocyte-platelet aggregates can plug such small vessels, resulting in either chronic conditions such as venous stasis ulceration or interfere with the effective radio- or chemotherapeutic treatment of solid tumors.

Computer simulations have begun to find success in bridging the gap in understanding between single cell behavior and large scale flows in the vasculature (King et al., 2001; King and Hammer, 2001b). Such theories model each molecular bond between cell and substrate as a linear spring with stochastic formation and breakage kinetics. The motion of cells suspended in a viscous fluid can be solved via boundary integral methods that are well suited to parallelization. Other investigators have developed techniques to study cell adhesion experimentally, by reconstituting the biological adhesion molecules in cell-free assays (Brunk et al., 1996). Selectin molecules, that mediate transient chemical adhesion between leukocytes, platelets, and endothelial cells physiologically, can be attached to a glass or polymer substrate, while carbohydrate selectin-binding ligand is attached to spherical polymer beads. The advantages of the cell-free assay are that a specific receptor-ligand pair can be isolated for study, with molecular densities on both surfaces precisely controlled. Flow chamber experiments with human leukocytes and endothelial cell monolayers have also been performed by several laboratories (e.g., Alon et al., 1995).

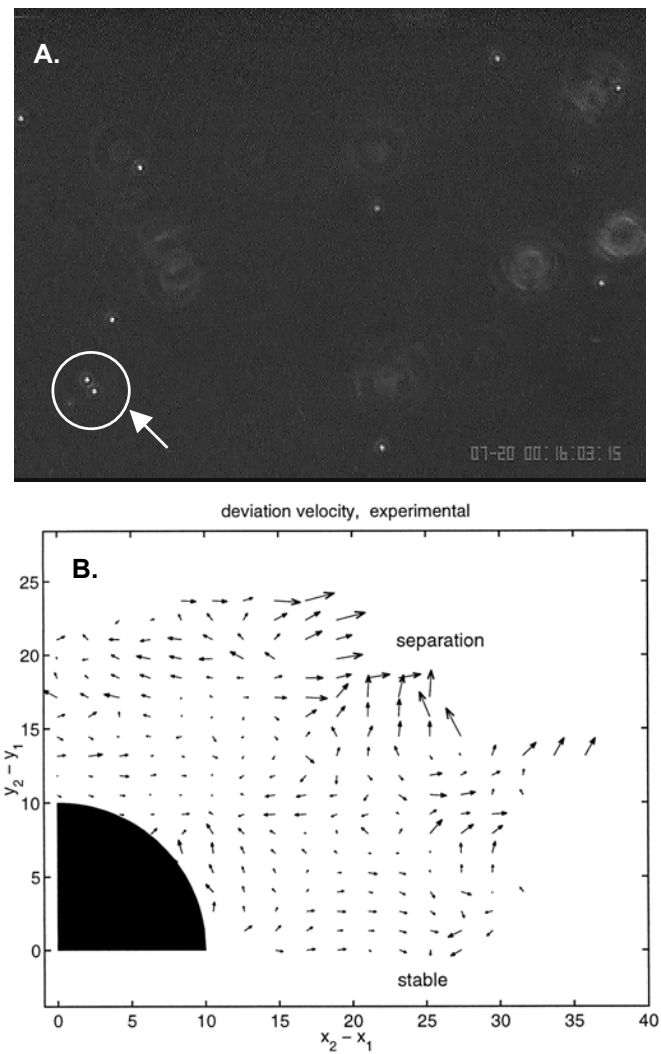

**Figure 2. A. Screen capture from in vitro experiment of sialyl Lewis(x)-coated beads rolling on P-selectin. A pair of beads is marked for binary analysis. B. Pairs of rolling beads (model cells) prefer certain center-to-center separation orientations over others. Deviation velocity map represents 209 pairs of beads.**

## METHODS

**Parallel-plate flow chamber**

Figure 1.A depicts a flow chamber to study the adhesion of cells to a functionalized surface, consisting of two parallel plates, the lower surface coated with a biological adhesion molecule. A parabolic pressure-driven flow is produced using a computer-controlled syringe pump. The cells or beads are slightly negatively buoyant, causing them to settle to the lower wall and interact adhesively. These interactions are detected by monitoring the translational velocity of the cells in the near-wall region from below using an inverted microscope. At the length scale of a single cell (radius a ~ 4 μm) the local flowfield near the wall can be approximated as a linear shear flow.

**Multiparticle adhesive dynamics**

We have developed a computational simulation of specific adhesive interactions between cells and surfaces under flow (Figure 1.B; King and Hammer, 2001a). In the adhesive dynamics formulation, adhesion molecules are modeled as compliant springs. One well-known model used to describe the kinetics of single biomolecular bond failure is due to Bell

$$k_r = k_r^0 \exp(r_0 F / k_b T)$$

which relates the rate of dissociation $k_r$ to the magnitude of the force on the bond $F$. Typical values for the unstressed off-rate $k_r^0$ and reactive compliance $r_0$ are 2 s$^{-1}$ and 0.04 nm for P-selectin binding with P-selectin glycoprotein ligand-1 (PSGL-1; Smith et al., 1999). The rate of formation directly follows from the Boltzmann distribution for affinity. The expression for the binding rate must also incorporate the effect of the relative motion of the two surfaces. The solution algorithm is as follows: (1) All unbound molecules in the contact area are tested for formation against the probability $P_f = 1 - \exp(-k_f \Delta t)$; (2) All of the currently bound molecules are tested for breakage against the probability $P_r = 1 - \exp(-k_r \Delta t)$; (3) The external forces and torques on each cell are summed; (4) The mobility calculation is performed to determine the rigid body motions of the cells; (5) Cell and bond positions are updated according to the kinematics of particle motion.

Unless firmly adhered to a surface, white blood cells can be effectively modeled as rigid spherical particles, as evidenced by the good agreement between bead versus cell in vitro experiments (Chang and Hammer, 2000). Typical values of physical parameters yield Reynolds numbers $\text{Re} = \dot\gamma a^2 = O(10^{-3})$, where $\dot\gamma = 100$ s$^{-1}$ is the shear rate, $a = 4$ μm is the cell radius, and $\nu = 1$ cS is the kinematic viscosity of the suspending fluid. Thus, inertia can be neglected and fluid motion is governed by the Stokes equation

$$\mu \nabla^2 u = \nabla p, \qquad \nabla \cdot u = 0$$

where $u$ is the velocity, $\mu$ is the fluid viscosity, and $p$ the local pressure. No-slip boundary conditions hold at the cell surfaces, and at $z = 0$, the position of the planar wall. The multiparticle problem is considerably more complex than the case of an isolated sphere, for which closed form solutions are available (e.g., Goldman, et al., 1967). We use a technique called the Completed Double-Layer Boundary Integral Equation Method (CDL-BIEM) (Phan-Thien et al., 1992). Applying the standard boundary element method to the Stokes flow problem produces a Fredholm integral equation of the first kind, which is generally ill-conditioned. By posing the mobility problem in terms of a compact double layer operator and completing the range with the velocity field resulting from a known distribution of point forces and torques placed inside each particle, one can derive a fixed point iteration scheme for solving the integral representation of the Stokes equation,

$$u(x) - u^\infty(x) = u^{RC}(x) + \oint_S \mathbf{K}(x-\xi)\cdot\phi(\xi)dS(\xi)$$

where $u^\infty$ is the ambient velocity in the absence of any particles, $u^{RC}$ is a "range completing" velocity generated by point forces and torques that accounts for the fact that the ill-behaved single layer integral has been discarded, $\mathbf{K}$ is the double layer operator, and $\phi$ is the unknown double layer distribution. After reducing the spectral radius of the corresponding discretization, the CDL-BIEM equation is found to converge rapidly. The presence of the wall is treated by incorporating the singularity solutions corresponding to a point force near a plane. The large separation of length scales between the deviation bond lengths and the particle radii requires very small time increments ($\Delta t = 10^{-7}$ s). To speed the calculation a coarse discretization is used that does not resolve particle-particle and particle-plane lubrication forces, which are added from known solutions as "external" forces.

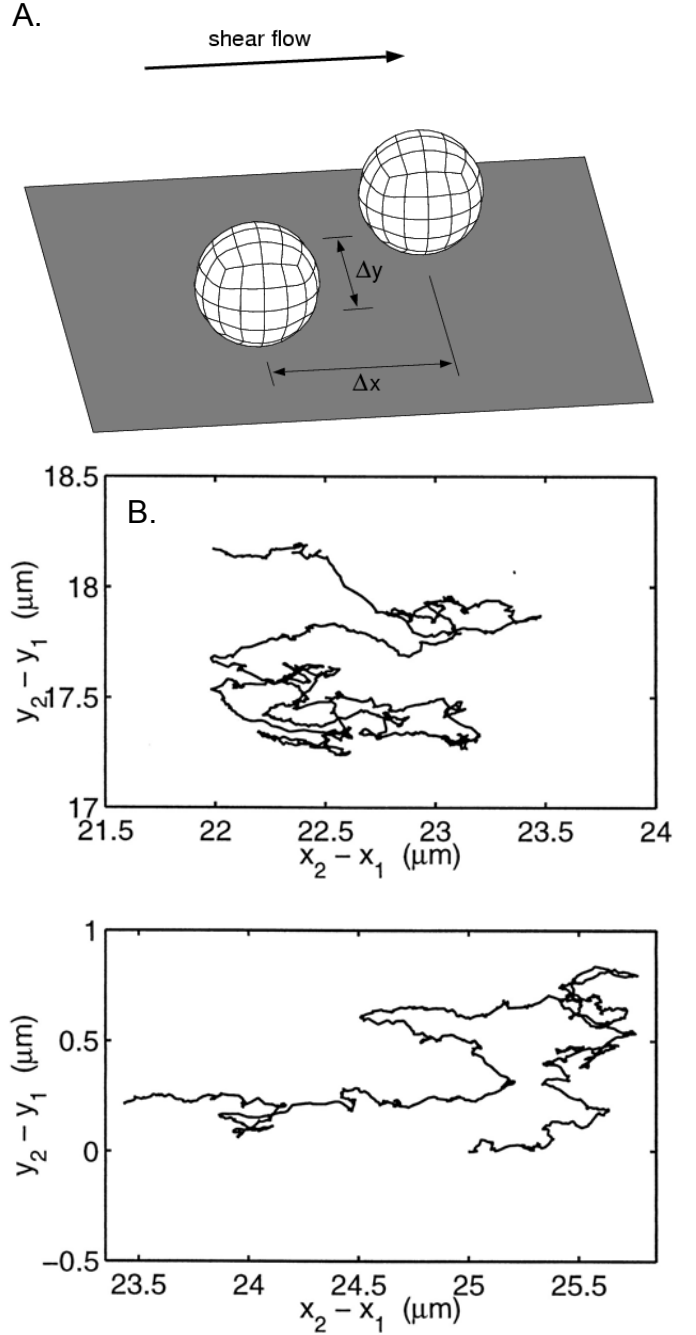

**Figure 3. A. In the MAD simulation, a nearby pair of rolling cells is approximated as two spheres above a plane. B. At high resolution, the simulation predicts that separation distance between two rolling cells undergoes a random walk in x-y space.**

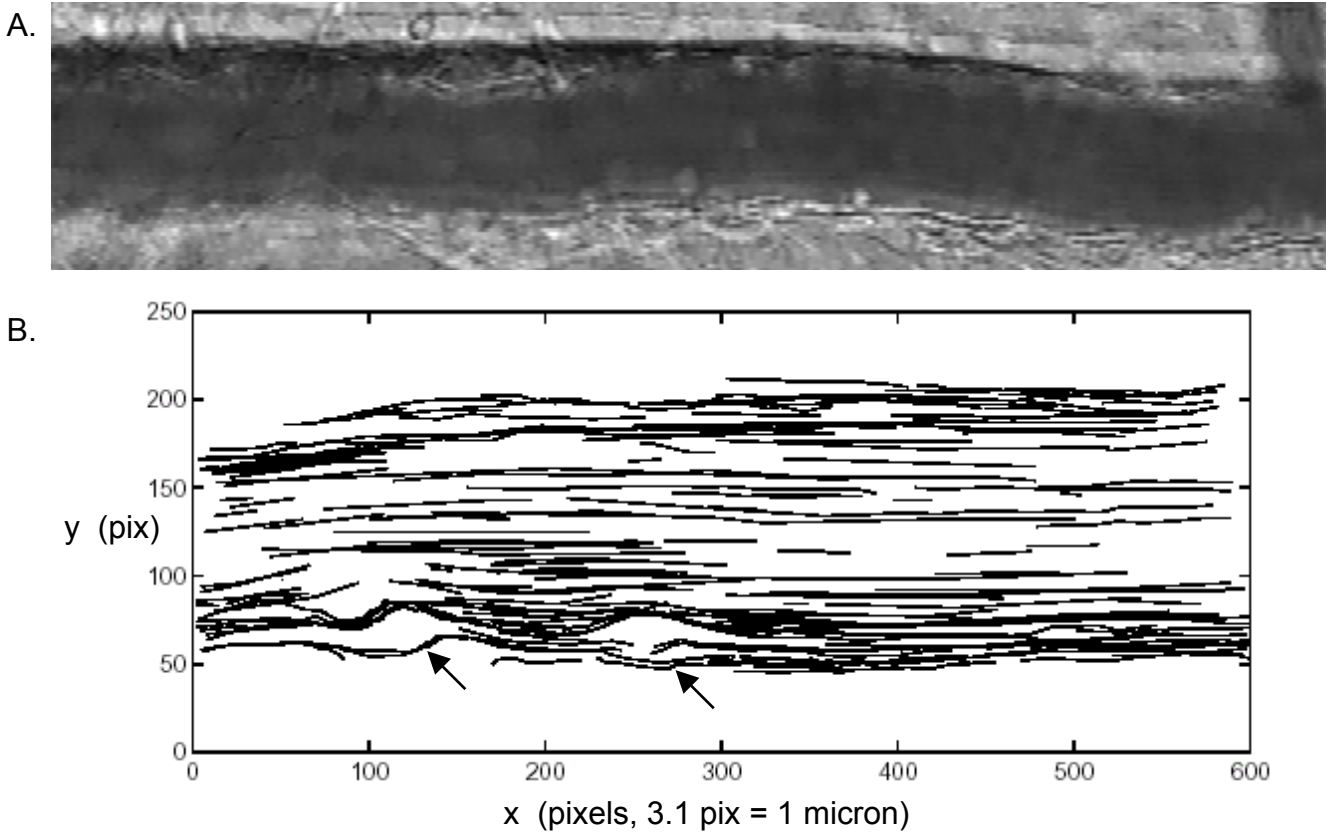

**Figure 4. In vivo microcirculatory flow experiments. A.** Digitized image of the cross-section of a 36 μm post-capillary venule in the cremaster muscle tissue of an anesthetized mouse. Several white blood cells interacting with the vessel wall are evident. **B.** Fluid streamlines yielded by tracking the motion of 0.5 μm fluorescent tracer beads in a 48 μm vessel similar to that pictured in A. Arrows show where the flow is diverted by adherent white blood cells at the vessel wall.

**Intravital microscopy**

We are fortunate to collaborate with Ingrid Sarelius (Pharmacology & Physiology, Univ. of Rochester), whose laboratory studies the inflammation cascade in live animal models. Male mice weighing 28 to 32g and older than 8 weeks are used in their experiments (Sarelius et al., 1981). Animals are anesthetized with an initial dose of sodium pentobarbital and maintained with supplemental doses as needed. Anesthetic level is monitored by observing withdrawal reflexes initiated by toe or tail pinch. Body temperature iss maintained by placing the animal on a glass heating coil connected to a water circulator set at 37°C. The animal is tracheotomized to establish a patent airway and two catheters placed in the right jugular vein for delivery of anesthetic and fluorescent flow tracer beads, respectively. The right cremaster muscle is exteriorized for *in situ* microscopy, producing a tissue section thin enough to allow light to pass through the sample (Baez, 1973; Sarelius et al., 1981). The tissue is superfused with a bicarbonate buffered salt solution throughout the surgical preparation and subsequent observation. The superfusate is maintained at physiological temperature (36±0.5°C) and equilibrated with gas containing 5% $CO_2$ and 95%$N_2$ to maintain physiological pH (7.40±0.05). At the end of each experiment, the animal is euthanized by i.v. injection of a lethal dose of anesthetic.

In this manner post-capillary venules can be observed using intravital confocal fluorescence microscopy. A microscope equipped with a water immersion objective, a Nipkow disk scanning confocal head, and a GenIII+ intensified CCD video camera are used to visualize fluorescent microspheres used as flow markers. For most venules, the confocal observation plane is focused at the center of the vessel lumen where the vessel walls outlining the observed cross-section are in clear focus. The thickness of the confocal slice has been calculated to be approximately 1 μm, defined as the distance where 50% fall off from source intensity occurred above and below a point source (Carlsson and Aslund, 1987). Flow patterns have been observed under control conditions, where slight inflammatory activation was present due to the surgical preparation of the tissue, and also under activated conditions induced by local injection of inflammatory cytokines prior to surgical preparation.

## RESULTS AND DISCUSSION

**Flow chamber experiments**

Figure 2.A shows a digitized image of an in vitro flow chamber experiment of sialyl Lewis(x)-coated beads (10 μm diameter) flowing over a P-selectin-coated surface at a shear rate of 100 $s^{-1}$. Sialyl Lewis(x) is the tetrasaccharide domain presented by many physiological ligands to the selectin molecules. Molecular densities on both surfaces were comparable to that found on wild-type cells, 90 sites/$\mu m^2$ on the beads and 180 sites/$\mu m^2$ on the lower wall of the flow chamber. As can be seen in Fig. 2.A, phase contrast microscopy produces images of good contrast, and cell/bead motions can be tracked via automated computer algorithms (e.g., using Matlab Imaging Toolbox or LabView). King and Hammer (2001a) studied the binary interactions between pairs of beads rolling in close vicinity, a comparable situation shown in Fig. 2.A. From a compilation of many such occurrences, the deviation motion of the rolling pair can be quantified (Figure 2.B). As shown in Figure 2.B, pairs of rolling beads will tend to separate when in certain configurations, however, the separation between the two beads will often evolve to a configuration where the beads are aligned with the flow direction in single-file. Such observations are consistent with previous in vitro studies of neutrophils (a subtype of white blood cells) where long linear trains of rolling cells have been observed to spontaneously form (Walcheck et al., 1996).

**Adhesive dynamics simulations**

The cell-surface interactions observed in flow chambers have been successfully recreated in detailed computer simulations. In particular, the interaction between pairs of cells rolling in close vicinity can be modeled as a true two-particle system (Figure 3.A). One advantage of the computer simulation is that behaviors can be examined at a much higher resolution (~1 nm) than is possible with optical microscopy. At this higher resolution it is apparent that the separation distance between two rolling cells undergoes a random walk in two-dimensions. From this random walk an apparent self-diffusivity in the plane can be defined. It has been shown that this diffusivity in the plane is decreased dramatically due to hydrodynamic cell-cell interactions for nearly-touching cells, as compared to widely-separated spheres (King and Hammer, 2001a).

**Intravital microscopy in live animal models**

An ultimate goal of the multiparticle adhesive dynamics simulation is to predict the dynamics of blood cell adhesion in the microcirculation, where the adhesion of white blood cells to the vessel wall is most important. Figure 4.A shows a digitized image of a real-time flow experiment in a cremaster muscle preparation in an anesthetized mouse. The post-capillary venule shown is about 36 μm in diameter, typical of vessels found between the smallest capillaries and larger collecting venules. From the surgical preparation, mildly inflammatory conditions result in the presentation of P-selectin adhesion molecule by the vessel wall and subsequent cell adhesion. The wall shear rate in such vessels has been previously characterized as between 50-400 $s^{-1}$ (Kim and Sarelius, in press). Tracking the motion of small fluorescent tracer beads introduced into the local microcirculatory network provides a means of visualizing the fluid streamlines (Figure 4.B). Previous investigators have used fluorescently-labeled red blood cells for this purpose, however, such larger tracers do not move with the local fluid velocity close to the wall and tend to collect at the vessel centerline due to well-understood mechanisms. Such flow measurements in intact vessels give the best estimates of the wall shear stress experienced by the endothelial cell layer lining the vessel interior, and will be useful in deriving improved estimates of mass transfer rates in this milieu of the circulation.

## CONCLUSIONS

Various in vitro, in vivo, and computational methods have been developed to understand the complex range of transient interactions between cells, neighboring cells, and bounding surfaces under flow. Knowledge gained from studying physiologically realistic flow systems may prove useful in microfluidic applications where the transport of blood cells and solubilized, bioactive molecules is needed, or in miniaturized diagnostic devices where cell mechanics or binding affinities can be correlated with clinical pathologies. Current work in our laboratory focuses on improving the MAD simulation to consider deformable cells of nonspherical (platelet) shape, cylindrical and complex branching geometries, and multiple receptor-ligand pairs, as well as the fabrication of glass capillary networks presenting biochemically realistic surfaces to flowing suspensions of cells.

## ACKNOWLEDGEMENTS

This work was funded by the National Institutes of Health, Grant No. HL18208. Figures 1.B, 2.B, and 3.B have been reproduced from King and Hammer (2001a), by copyright permission of the Biophysical Society.